\documentclass[
reprint,
onecolumn,
 amsmath,amssymb,
 aps,
]{revtex4-2}

\usepackage{graphicx}
\usepackage{dcolumn}
\usepackage{bm}
\usepackage{xcolor}

\begin{document}

\preprint{APS/123-QED}

\title{On the rheoscopic measurement of turbulent decay in wall-bounded flows}

\author{Tao Liu$^{1,2}$}
\email{tao.liuuga@hotmail.com}
\author{Victoria Nicolazo-Crach$^{1,3}$}
\author{Ramiro Godoy-Diana$^{1}$}
\author{Jos\'{e} Eduardo Wesfreid$^{1}$}
\author{Beno\^{i}t Semin$^{1}$}

\affiliation{$^{1}$PMMH, CNRS, ESPCI Paris, Universit\'{e} PSL, Sorbonne Universit\'{e}, Universit\'{e} Paris Cit\'{e}, Paris, 75005, France}
\affiliation{$^{2}$Department of Earth, Planetary, and Space Sciences, University of California, Los Angeles, CA 90095, USA}
\affiliation{$^{3}$FAST, CNRS, Universit\'{e} Paris-Saclay, 91405 Orsay, France}

\date{\today}

\begin{abstract}
Quench experiments where the flow passes from a fully turbulent state to a laminar state by an
abrupt decrease in the flow Reynolds number ($Re$) have been extensively studied in the literature
to quantify the turbulent-laminar transition process in wall-bounded flows. Measurements have been classically made using rheoscopic fluid visualisations, which
make turbulent coherent structures easily identifiable, allowing for quantification of the evolution
of a turbulent fraction—the percentage of a given observation window where turbulence is deemed
active by the presence of coherent structures, such as  streamwise vortices called rolls, and modulations of the streamwise velocity fluctuations called streaks. Decay characteristic times of these structures have therefore been extensively measured. However, owing to the nature of visualization based techniques, only a single decay time is typically extracted, whereas measurements of the velocity field can reveal distinct decay times associated with different velocity or kinetic energy components. As a result, the physical meaning of the decay time inferred from visualization alone is not straightforward. The goal of the present paper is to perform such a comparison quantitatively,
using particle image velocimetry (PIV) measurements and rheoscopic fluid visualisations in the same
setup: a Couette-Poiseuille experiment. We observe via PIV different characteristic times of decay
for streamwise (streaks) and spanwise (rolls) velocity fluctuations. We show that the characteristic
time of decay of the turbulent fraction observed by visualisation is close to the decay of the
streaks.

\end{abstract}

\maketitle


\section{Introduction}
\label{sec:intro}

Rheoscopic visualization is widely used to identify coherent structures in wall-bounded shear flows, including spiral bands in Taylor–Couette flow \cite{Coles_1965,Lerma_1985}, turbulent spots in plane Couette flow \cite{Tillmark_Alfredsson_1992}, turbulent regions in plane Poiseuille flow \cite{Carlson_1982, Lemoult_2012}, and oblique bands in plane Poiseuille and Couette–Poiseuille flows \cite{Paranjape_Duguet_Hof_2020,Klotz_Wesfreid_2017}.
The method relies on seeding the flow with small, weakly settling, anisotropic oblate flakes that remain suspended for long durations \citep{Matisse_1984}.
In simple parallel shear flows, such particles undergo Jeffery type rotations and align on average along preferred orientations \citep{Jeffery_1922}, a behavior that Savaş \citep{Savaş_1985} interpreted as alignment with the local stream surface.

This interpretation does not apply to general three-dimensional flows.
Analyses of three-dimensional velocity fields \citep{Gauthier_1998,Hecht_2010} have shown that the orientation of anisotropic flakes is controlled by the eigenstructure of the local velocity-gradient tensor rather than by stream surfaces.
As a result, rheoscopic images preferentially highlight regions dominated by shear, such as the peripheries of Taylor vortices and the boundary layers in rotating-disk flows \cite{Gauthier_1998}.
Direct comparisons with PIV \cite{Abcha_2008} have demonstrated that the reflected intensity correlates strongly with radial and shear motions in Taylor–Couette flow.
The experimental and numerical study of flow in a precessing sphere \cite{Goto_Kida_Fujiwara_2011} further showed that the observable intensity originates from the non-uniform orientation of flakes rather than from spatial accumulation and that flake normals evolve in the same way as infinitesimal material surface elements advected by the flow.
Recent Fokker–Planck formulations describe the intensity field as the Eulerian probability distribution of flake orientations shaped by velocity gradients, advection, and rotational diffusion \cite{Itano_2025}.

Although these studies have advanced our understanding of rheoscopic visualization, the quantitative relationship between visualization in wall-bounded shear flows and the underlying velocity field remains insufficiently explored. Visualization has been used to infer turbulence lifetimes in plane Couette flow \cite{Bottin_1998_EPJ} and to define order parameters such as the turbulent fraction, which is the area identified as turbulent through image processing, in studies of directed percolation in plane Poiseuille \cite{Sano_NP_2016} and plane Couette \cite{Lemoult_2016} flows. These studies often employ quench experiments, in which the flow is abruptly reduced from a fully turbulent state to a laminar or transitional regime \cite{Bottin_1998_EPJ,Souza_JFM_2020,Cerbus_Mullin_2024}.

The quenching events reveal distinct decay dynamics of the coherent structures, namely streaks and rolls, that sustain turbulence in wall-bounded shear flows. Here, streaks are defined as elongated streamwise-velocity modulations, whereas rolls are streamwise-oriented vortices that generate streaks via the lift-up mechanism \citep{Landahl_1975, Liu_etal_2021, Gome_2024}. Particle image velocimetry allows these structures to be identified directly through their associated velocity components; however, rheoscopic visualizations provide only indirect signatures of the flow organization. The correspondence between the structures identified through visualization, such as streaks or rolls, and those present in the velocity field, including their respective decay dynamics, is not fully established. These open questions motivate a direct, quantitative assessment of how rheoscopic visualization relates to the underlying velocity field.

We therefore perform controlled quench experiments in plane Couette-–Poiseuille flow and carry out a systematical comparison between rheoscopic visualization and PIV measurements in the streamwise–spanwise plane.
Image-processing methods inspired by earlier studies \cite{Bottin_thesis_1998,Sano_NP_2016} are used to extract the turbulent regions from visualization, and the results are compared with velocity thresholding criteria obtained from PIV.
Our aim is to quantify the agreement between the two diagnostics, identify systematic differences in their temporal evolution, and determine the decay time statistics.

The article is organized as follows: Section~\ref{sec:set-up} details the experimental set-up, diagnostics, and protocols; Section~\ref{sec:methods} presents the image processing methods for detecting turbulent regions; Section~\ref{sec:results} analyzes the turbulent fraction and decay time results; and Section~\ref{sec:Conclusion} provides the discussion and conclusions.

\section{Experimental set-up and measurements technique}
\label{sec:set-up}

\subsection{Experimental set-up}
\label{subsec:Setup}

\begin{figure}[t!]
  \centerline{\includegraphics[width=15cm]{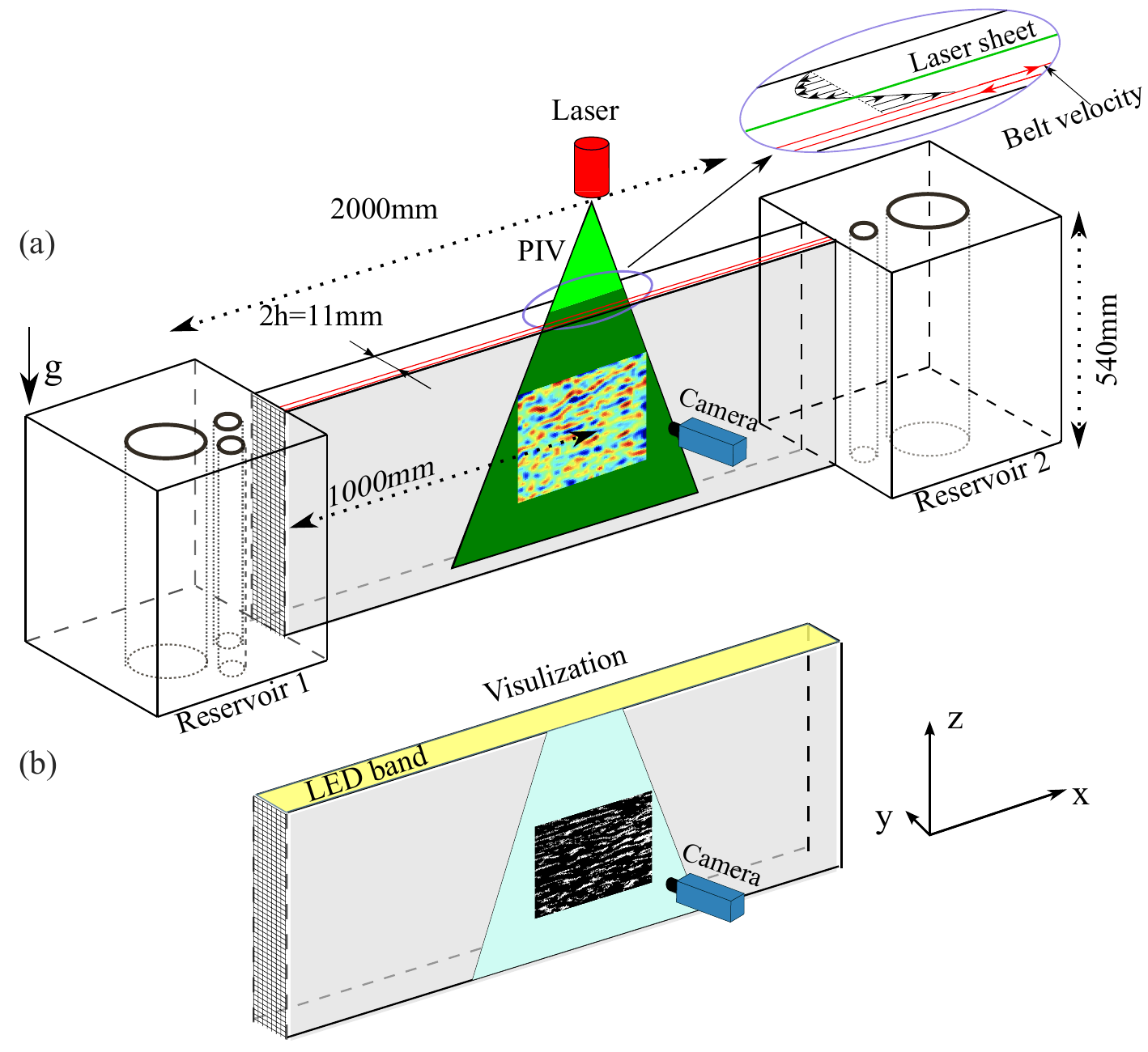}}
  \caption{Schematic of the plane Couette–Poiseuille flow channel equipped for (a) 2D2C Particle Image Velocimetry (PIV) and (b) flow visualization measurements with LED light band and anisotropic light-reflecting aluminium particles. }
\label{fig:setup}
\end{figure}

A schematic of the apparatus is shown in Fig.~\ref{fig:setup}. The experiment was performed in a plane Couette–Poiseuille channel driven by a translating Mylar belt, following the configuration described by Liu \emph{et al.} (2021). Two parallel glass plates, separated by a constant gap of width \(2h = 11.0 \pm 0.3~\mathrm{mm}\), form the sidewalls of the test section. The channel connects two water reservoirs located at its upper and lower ends. The Mylar belt is guided around vertical cylinders in the reservoirs such that one side of the belt moves parallel to one glass plate, acting as a moving wall.

The belt is driven by a servomotor whose angular velocity is controlled via a \textsc{Labview} interface. The belt motion generates a Couette shear flow, which simultaneously induces a hydrostatic pressure difference between the two reservoirs. The resulting counterflow produces a Poiseuille component, such that the net mean flux through the channel is nearly zero. The characteristic Reynolds number is defined by
\begin{equation}
Re = \frac{U_{\mathrm{belt}} h}{\nu},
\end{equation}
where \(U_{\mathrm{belt}}\) is the belt velocity and \(\nu\) the kinematic viscosity of water, evaluated from the measured temperature \(T = 23.1 \pm 0.4^{\circ}\mathrm{C}\).
The channel has a streamwise length $L_x = 2000~\mathrm{mm}$ and a spanwise width $L_z = 540~\mathrm{mm}$, corresponding to aspect ratios or nondimensional lengths $x^* = L_x/h = 364$ and $z^* = L_z/h = 98$.

\subsection{Particle image velocimetry}
\label{subsec:PIVTech}

Two-dimensional particle image velocimetry (PIV) was performed in the streamwise--spanwise (\(x\)--\(z\)) plane. The fluid was seeded with polyamide particles, \(20~\mu\text{m}\) in diameter and with a density of \(1.03~\mathrm{g~cm^{-3}}\). The polyamide particle concentration in water was therefore approximately \( C_{\text{PIV}} \sim 1.10 \times 10^{-5}~\mathrm{g~mL^{-1}} \), which is well below the Einstein limit for significant viscosity modification. Illumination was provided by a dual-pulse Nd:YLF laser 
(Darwin~Duo, 527~nm, 20~mJ per pulse), forming a light sheet of 
thickness about 1~mm at the zero laminar velocity location 
\(y/h = 0.33 \pm 0.04\). The inter-pulse delay was set to 
\(\Delta t = 12~\mathrm{ms}\), which produced a particle 
displacement of a few pixels between frames.
The laser, camera, and motor were synchronised using a 
LaVision\textsuperscript{\textregistered} programmable timing unit.

Image pairs were processed with DaVis~10 using a 
multi-pass cross-correlation algorithm with elliptical interrogation 
windows (aspect ratio 4:1, total area 2304~pixels) and 50\,\% overlap. 
Spurious vectors near image boundaries were removed, and the final 
velocity field was cropped to \(65h \times 67h\).

\subsection{Flow visualisation}
\label{subsec:FlowVisuTech}

Flow visualization was performed using a rheoscopic suspension of aluminium flakes (Stapa Il Hydralan 2154; diameter 18--24~\textmu m) dispersed in water at a concentration of \(C_\mathrm{Al} \simeq 4\times10^{-6}~\mathrm{g\,ml^{-1}}\). This concentration is well below the Einstein limit for significant viscosity modification. The suspension was illuminated by LED light bands placed atop the channel (see Fig.~\ref{fig:setup}(b)). 
The suspension was homogenized by circulating the belt for several minutes 
prior to each experiment to ensure a uniform particle distribution. 

Images were acquired with a LaVision\textsuperscript{\textregistered} Imager~MX5M camera at a 
resolution of 1908~px~$\times$~2176~px, fitted with a Nikon\textsuperscript{\textregistered} 
17--35~mm~f/2.8 lens and operated at 1~Hz in continuous mode. 
A calibration target provided a spatial scale of 
\(0.20~\mathrm{mm\,px^{-1}}\), corresponding to a field of view of 
approximately \(380\times435~\mathrm{mm^2}\) 
(\(69h\times79h\)). The camera was positioned normal to the glass plate 
at a distance of 0.97~m. The optical configuration was identical to that 
used for the PIV measurements to facilitate direct quantitative 
comparison between the two diagnostics.
\subsection{Quench protocol}
\label{subsec:QuenchProtocol}

Each experiment began with the flow fully turbulent at an initial Reynolds number \(Re_i = 1000\).
At \(t = 0\), the belt velocity was abruptly reduced to a lower final Reynolds number \(Re_f\).
The deceleration time was less than \(0.1~\mathrm{s}\), corresponding to less than two advective time units \(h/U_{\mathrm{belt}}\), and therefore negligible compared with the typical turbulence decay time.
For each \(Re_f\), three realisations were recorded for both flow visualisation and PIV, respectively.
This procedure allowed for direct comparison of the temporal evolution of the turbulent fraction and the kinetic energy decay obtained from the two complementary diagnostics under strictly identical experimental conditions.

\section{Methods for Detecting Turbulent Regions}
\label{sec:methods}
In this section, we describe the image-processing procedure applied to the flow visualization data to identify turbulent regions. The detection method follows the approaches developed in previous studies of the directed percolation in plane Poiseuille flow \cite{Sano_NP_2016} and turbulence lifetime in plane Couette flow \cite{Bottin_thesis_1998}.

\subsection{Detection of Turbulent Regions by Image Intensity Variation}
\label{subsec:MethodSano}

Turbulent regions generate markedly stronger intensity fluctuations than laminar flow \cite{Dauchot_Daviaud_1995}. Building on this contrast, \cite{Sano_NP_2016} introduced an image-processing approach that reliably detects turbulent patches in transitional plane Poiseuille flow. In the present study, we employ a modified version of their method by adding an image intensity normalization step to improve the contrast between active turbulent and laminar regions when processing flow visualization images and quantifying the decay of turbulence in plane Couette Poiseuille flow experiments.

A laminar reference state was constructed from 100 visualization images recorded at
$Re_f = 300$, for which the flow is fully laminar and free of streak structures. For each
pixel $(x,z)$, we compute the temporal mean intensity $I_{\mathrm{lam}}(x,z)$.

\begin{figure}[b!]
  \centerline{\includegraphics[width=15cm]{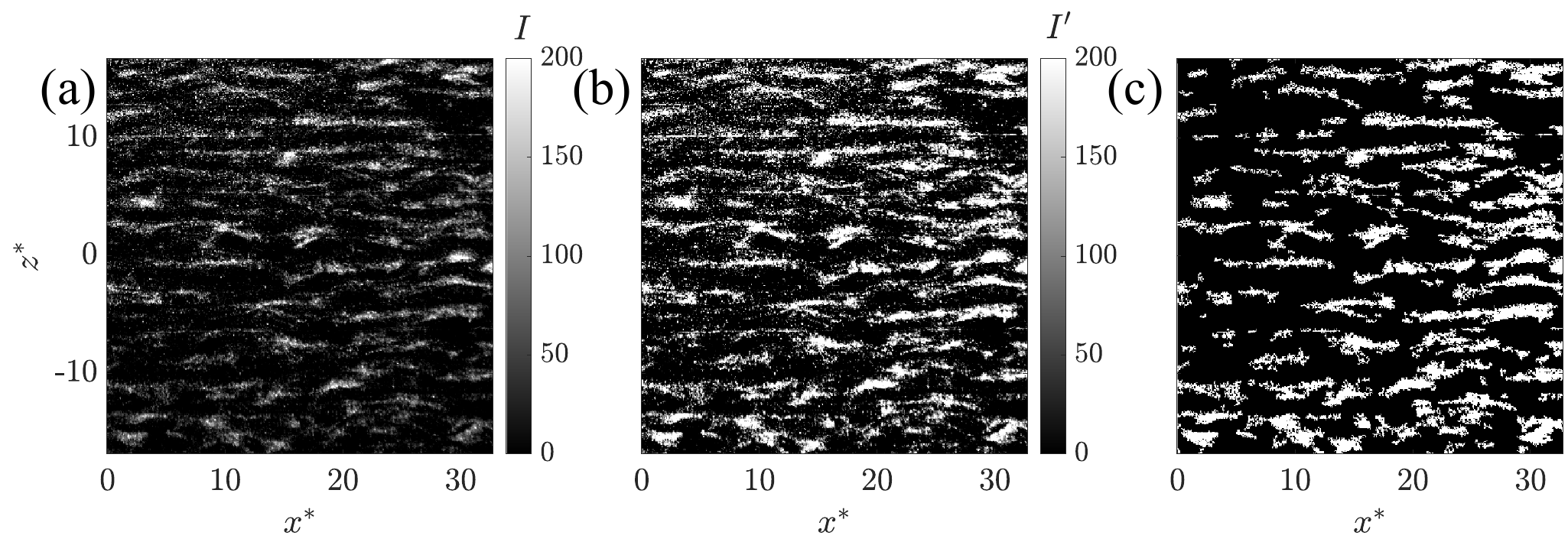}}
  \caption{Illustration of image processing steps proposed by \cite{Sano_NP_2016}  for a sample flow visualization image taken shortly after the quench at \( t^* = 62 \) for \( Re_f = 500 \): (a) Raw grayscale image, (b) intensity-normalized image to enhance contrast, (c) binary image obtained via intensity fluctuation thresholding.}
\label{fig:image_process_Sano}
\end{figure}

From the mean laminar field $I_{\mathrm{lam}}(x,z)$, we extract the global minimum and
maximum intensities, denoted $I_{\min}$ and $I_{\max}$, respectively. These values are used to
normalize all visualization frames into an 8-bit range in order to increase contrast (see Fig.~\ref{fig:image_process_Sano}(a,b)),
\begin{equation}
    I^{\ast}(x,z,t)
    = \frac{\, I(x,z,t) - I_{\min} \,}{I_{\max} - I_{\min}} \times 255.
\end{equation}

Each normalized image is then background-subtracted using the normalized
laminar reference field $I_{\mathrm{lam}}^{\ast}(x,z)$ to correct for spatial light inhomogeneity:
\begin{equation}
    I'(x,z,t) = I^{\ast}(x,z,t) - I_{\mathrm{lam}}^{\ast}(x,z).
\end{equation}

We then compute the temporal standard deviation $\sigma^{\ast}_{\mathrm{lam}}(x,z)$ of the
normalized laminar background images. Following the criterion introduced by
\cite{Sano_NP_2016}, the deviation of each instantaneous image,
$I'(x,z,t)$, is compared against the threshold $n\,\sigma^{\ast}_{\mathrm{lam}}(x,z)$
for each pixel, with $n = 3$ as a conventional choice. Pixels satisfying
\begin{equation}
    |I'(x,z,t)| > 3\,\sigma^{\ast}_{\mathrm{lam}}(x,z)
\end{equation}
are classified as turbulent. The thresholded image is shown in binary form in Fig.~\ref{fig:image_process_Sano}(c), and they provide a clear identification of the turbulent structures, such as the streamwise streaks.

\subsection{Detection of Turbulent Regions by Image Morphological Method}
\label{subsec:MethodBottin}

\begin{figure}[t!]
  \centerline{\includegraphics[width=11cm]{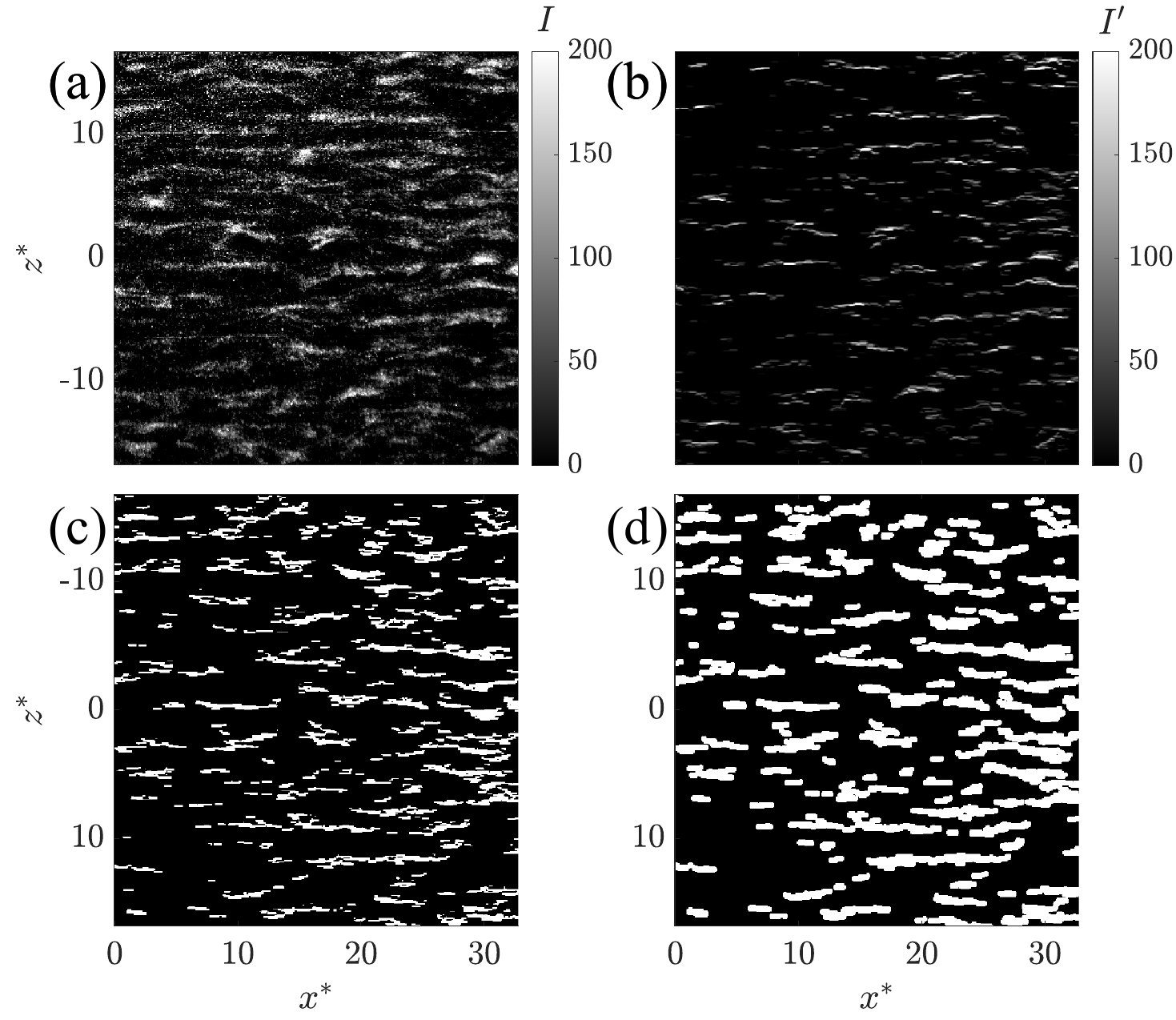}}
  \caption{Illustration of the image-processing procedure following \cite{Bottin_thesis_1998} applied to a sample flow-visualization frame taken shortly after the quench at \( t^* = 62 \) for \( Re_f = 500 \): (a) Raw grayscale image; (b) morphological opening followed by gradient filtering; (c) thresholded binary field; and (d) image erosion and dilation.}
\label{fig:image_process_Bottin}
\end{figure}

Fig.~\ref{fig:image_process_Bottin} shows the image-processing procedure
applied to the flow-visualization frames, following the method originally
developed for plane Couette flow to study turbulence decay statistics
\cite{Bottin_thesis_1998}. 

A horizontal morphological opening is performed using a \(7 \times 7\) structuring element:

\[
\begin{bmatrix}
0 & 0 & 0 & 0 & 0 & 0 & 0 \\
0 & 0 & 0 & 0 & 0 & 0 & 0 \\
1 & 1 & 1 & 1 & 1 & 1 & 1 \\
1 & 1 & 1 & 1 & 1 & 1 & 1 \\
1 & 1 & 1 & 1 & 1 & 1 & 1 \\
0 & 0 & 0 & 0 & 0 & 0 & 0 \\
0 & 0 & 0 & 0 & 0 & 0 & 0 \\
\end{bmatrix}.
\]
This kernel acts as a spanwise size filter, retaining only structures whose spanwise (\(z\)-direction) extent exceeds 3 pixels (corresponding to approximately \(0.2h\)). By imposing this minimum spanwise width, the operation suppresses fine-scale noise and short-wavelength spanwise fluctuations while preserving the dominant elongated streak patterns aligned in the streamwise (\(x\)) direction. The structuring element size was chosen based on the typical streak width observed in visualizations and the spatial resolution of the imaging system.

To correct illumination nonuniformity, a $3 \times 3$ north-south gradient filter,
\[
\begin{bmatrix}
-1 & -1 & -1 \\
\;\,0 & \;\,0 & \;\,0 \\
\;\,1 & \;\,1 & \;\,1
\end{bmatrix},
\]
is subsequently applied. This step removes spanwise variations caused by
top- or bottom-mounted light sources, which commonly introduce inhomogeneous
intensity gradients in flow-visualization images. The raw image, shown in Fig.~\ref{fig:image_process_Bottin}(a), has intensity \(I\). After applying opening and gradient filters, the processed image with intensity \(I'\) is presented in Fig.~\ref{fig:image_process_Bottin}(b).

After thresholding the image using the pixel-wise mean intensity obtained from 100 fully turbulent reference frames (e.g.\ at $Re_i = 1000$), the resulting binary turbulent field is shown in Fig.~\ref{fig:image_process_Bottin}(c). To refine this field, morphological erosion followed by dilation is applied to remove isolated noisy pixels, break spurious thin connections, and smooth the boundaries of the turbulent patches. This yields a clean, coherent binary mask that reliably captures dynamically active regions as revealed by the visualization.

\section{Results}
\label{sec:results}


\subsection{Turbulent fraction evolution}

The turbulent fraction $F$ is defined as the ratio of the area occupied by turbulent regions to the total measurement area:
\begin{equation}
F = \frac{A_{\mathrm{turb}}}{A_{\mathrm{tot}}},
\end{equation}
where $A_{\mathrm{turb}}$ denotes the area identified as turbulent and $A_{\mathrm{tot}}$ is the total measurement area. Turbulent regions are detected using the image processing methods described in Section~\ref{sec:methods}. For comparison across datasets, we normalize $F$ by its mean value in the initial fully turbulent state at $Re_i = 1000$, such that $F(t \le 0) \approx 1$.
\begin{figure}[b!]
  \centerline{\includegraphics[width=15cm]{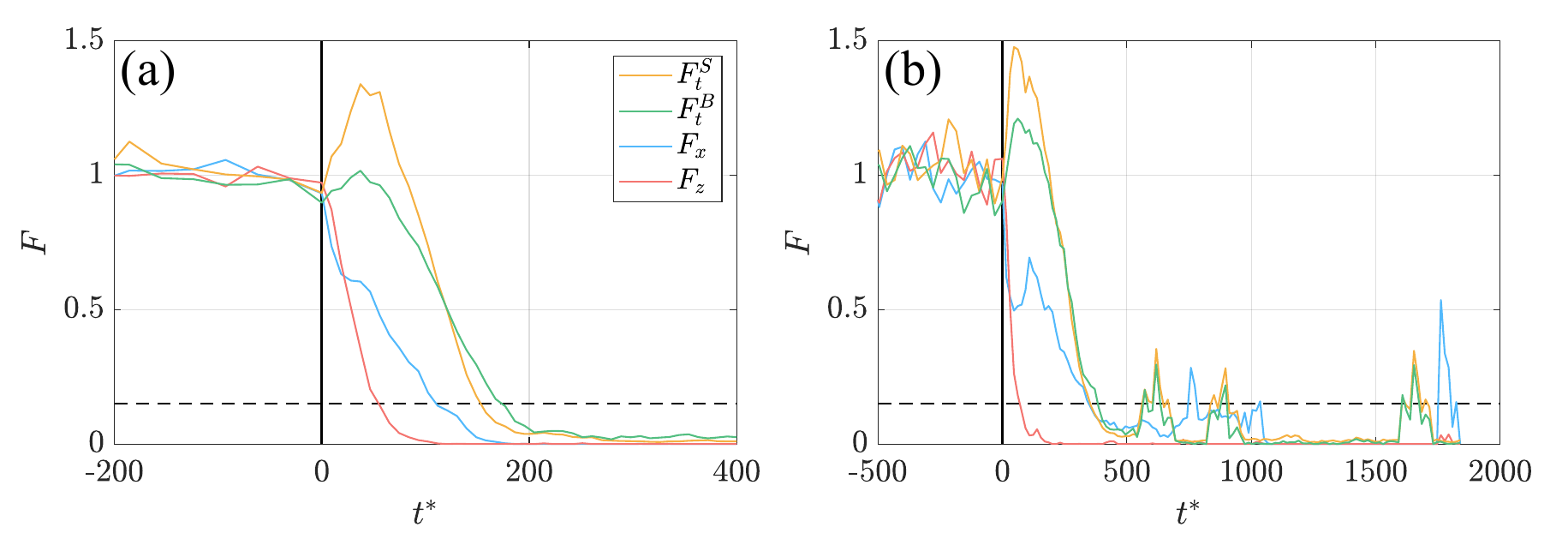}}
  \caption{Comparison of the temporal evolution of the turbulent fraction of quench experiments obtained flow visualization and PIV velocity field results. (a) $Re_f = 300$, (b) $Re_f = 500$. The gold solid line $F_t^S$ represents flow visualization analyzed using the image processing method in Section~\ref{subsec:MethodSano}. The green solid line $F_t^B$ shows an alternative flow visualization result processed following the method in Section~\ref{subsec:MethodBottin}. The blue solid line corresponds to $F_x$, obtained by thresholding the streamwise velocity field with $|u_x| > 0.1U_{\text{belt}}$, while the red solid line corresponds to $F_z$, obtained by thresholding the spanwise velocity field with $|u_z| > 0.05U_{\text{belt}}$. The horizontal black dashed line indicates the reference level $F = 0.15$.}
\label{fig:turb_frac_evolution}
\end{figure}


Fig.~\ref{fig:turb_frac_evolution}(a,b) shows the temporal evolution of $F$ following the quench to $Re_f = 300$ and $Re_f = 500$, respectively. For $Re_f = 300$, the turbulent fractions obtained from image-based detection, $F_t^S$ (yellow, intensity-variation method) and $F_t^B$ (green, morphological method), both exhibit a small peak shortly after the quench before subsequently decaying. The peak is more pronounced in $F_t^S$, while $F_t^B$ displays a weaker overshoot.

The velocity-based turbulent fraction $F_x$, defined by the criterion $|u_x| > 0.1 U_{\text{belt}}$, decays nearly monotonically but shows a brief plateau that suggests a slight peak. This behavior is consistent with the transient lift-up effect produced by streamwise vortices \citep{Landahl_1975}. A similar stronger peak also appears in $F_x$ at $Re_f = 500$, occurring during the initial stages of the transient decay.


It is worth noting that both image-based turbulent fractions correspond to the same experimental realization, whereas the velocity-threshold results come from a separate PIV realization. However, at these transitional Reynolds numbers the quenching dynamics are highly repeatable, as demonstrated previously (Fig.~13 in \cite{Liu_etal_2021}).

\begin{figure}[b!]
  \centerline{\includegraphics[width=15cm]{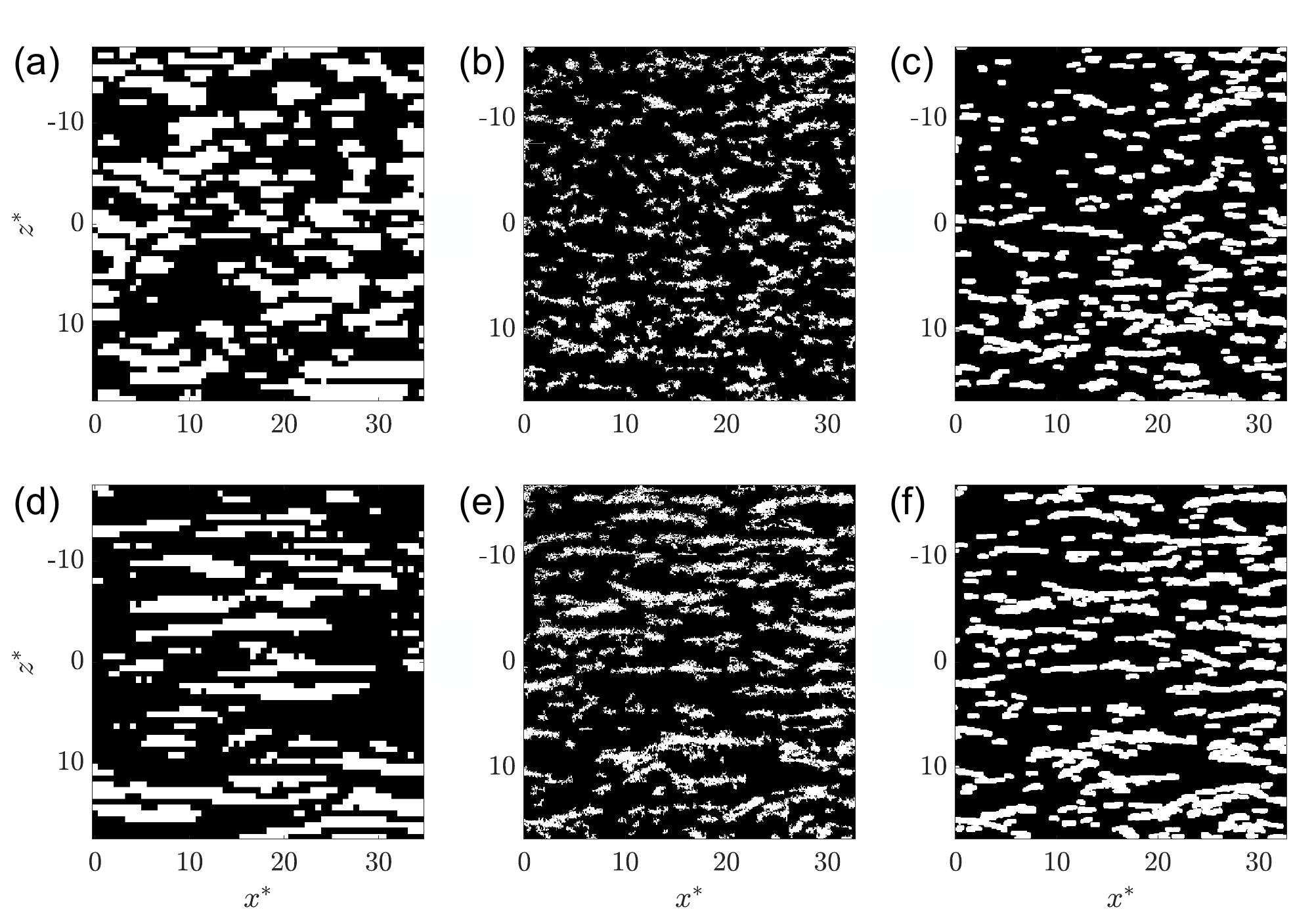}}
  \caption{
Binary flow fields for the quench case at $Re_f = 500$. 
Top row: fully turbulent initial state at $Re_i = 1000$ ($t<0$ in Fig.~\ref{fig:turb_frac_evolution}b). 
Bottom row: post-quench state when the turbulent fraction reaches its peak. 
Panels (a,d) show velocity-thresholded fields ($\lvert u_x\rvert \ge 0.1\,U_{\mathrm{belt}}$) from PIV measurements, 
(b,e) intensity-variation thresholding, and (c,f) image morphological processing.
}
\label{fig:turb_trans_comp}
\end{figure}

To clarify the physical origin of the peaks in $F_t$ and $F_x$, we compare binary turbulent-field snapshots in Fig.~\ref{fig:turb_trans_comp}. The fully turbulent state is shown in panels~(a--c), characterized by fragments of various sizes corresponding to a broad spectrum of waves or undulations that break up the streaks. The corresponding flow fields at peak intensity after quenching are displayed in panels~(d--f), with panel~(d) corresponding to \( t^* = 108 \) and panels~(e) and~(f) to \( t^* = 62 \).

In the velocity-thresholding results (panels a and d), the peak coincides with a transient elongation and straightening of turbulent streaks during relaminarization, while the turbulent area in the fully turbulent state is significantly larger than that at the peak. In the flow visualizations (panels b, c vs.\ e, f), the peak similarly corresponds to a transient elongation (straight streaks) of turbulent streaks during relaminarization.

In the fully turbulent regime, turbulent streaks appear highly fragmented, with a prevalence of differently sized fragments that correspond to a wide spectrum of small-scale waves or undulations breaking the streaks. In contrast, after the quench, the streaks evolve into longer and wider coherent structures. The remaining fragments are more coherent and exhibit a narrower size distribution. These correspond to well-defined undulations that temporarily induce normal vorticity, thereby regenerating the rolls and streaks as predicted by the SSP scenario. This coherent growth temporarily increases the detected turbulent area, giving rise to the observed transient peak. Such overshoots in the turbulent fraction have been reported previously in plane Couette flow experiments \citep{Bottin_1998_EPJ,Bottin_thesis_1998} and have recently been described as secondary peaks in numerical simulations of plane Couette-Poiseuille flow \citep{Etchevest2025}.

Aside from the transient peak, the temporal evolutions of 
$F_t^S$ and $F_t^B$ at $Re_f = 300$ decay on a timescale that is 
slightly longer, although still comparable, to that of $F_x$.  
This can be seen from the dashed line indicating the level 
$F = 0.15$, which is the criterion used in \cite{Liu_etal_2021}. We define the decay as the time when the turbulent fraction drops to 0.15 of its initial value. This threshold choice has been justified in previous work \cite{Liu_etal_2021}. All three measures exhibit decay times that are significantly longer 
than that obtained from the spanwise–velocity threshold $F_z$ using 
$|u_z| > 0.05\,U_{\text{belt}}$.  
In contrast, at $Re_f = 500$, the decay times of $F_t^S$, $F_t^B$, and 
$F_x$ collapse onto nearly the same value.

\begin{figure}[t!]
  \centerline{\includegraphics[width=15cm]{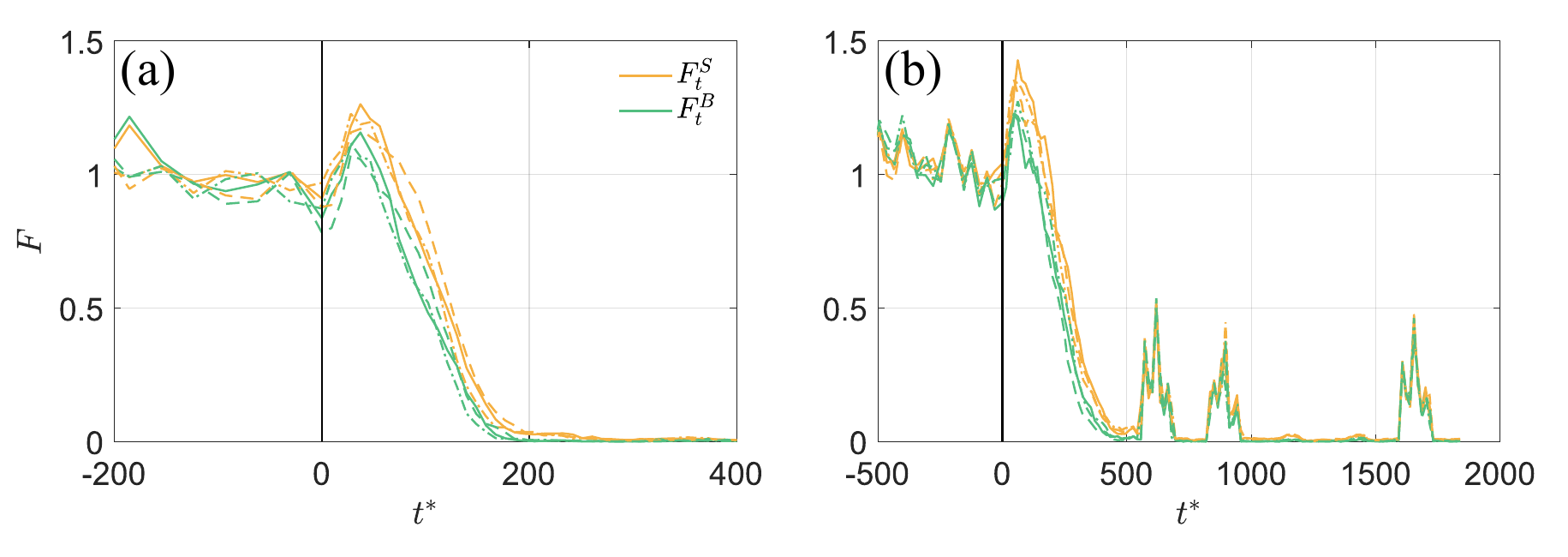}}
  \caption{Temporal evolution of the turbulent fraction in quench experiments, extracted from flow visualization using the intensity-variation method \cite{Sano_NP_2016} (gold lines) and the morphological method \cite{Bottin_thesis_1998} (green lines). Panels show three repeated runs at (a) $Re_f=300$ and (b) $Re_f=500$, with different line types indicating individual realizations.}
\label{fig:turb_frac_evolution_repet}
\end{figure}
The discrepancy at the lower final Reynolds number likely reflects the 
fact that, during the final stages of relaminarisation, the residual streaks 
have extremely weak velocity amplitudes and therefore fall below the 
velocity-threshold criterion.  
However, the image-processing methods remain sensitive to the weak 
shear \cite{Weidman_1976,Carlson_1982} associated with these fading streaks, allowing them to track the streak for a longer time.

To evaluate the robustness of the visualization-based detection methods,
we applied them to three independent experimental realizations at each
final Reynolds number.  
Fig.~\ref{fig:turb_frac_evolution_repet}(a) and (b) show the resulting
turbulent-fraction evolutions for $Re_f = 300$ and $Re_f = 500$, with
different line styles corresponding to the separate runs.  
Although small run-to-run variations are present, the overall temporal
trends and decay times are highly consistent across realizations,
indicating good reproducibility of the experiments and robust performance
of both image-processing methods.

\subsection{Decay time}
\label{subsec:decay_time}
\begin{figure}[b!]
  \centerline{\includegraphics[width=10cm]{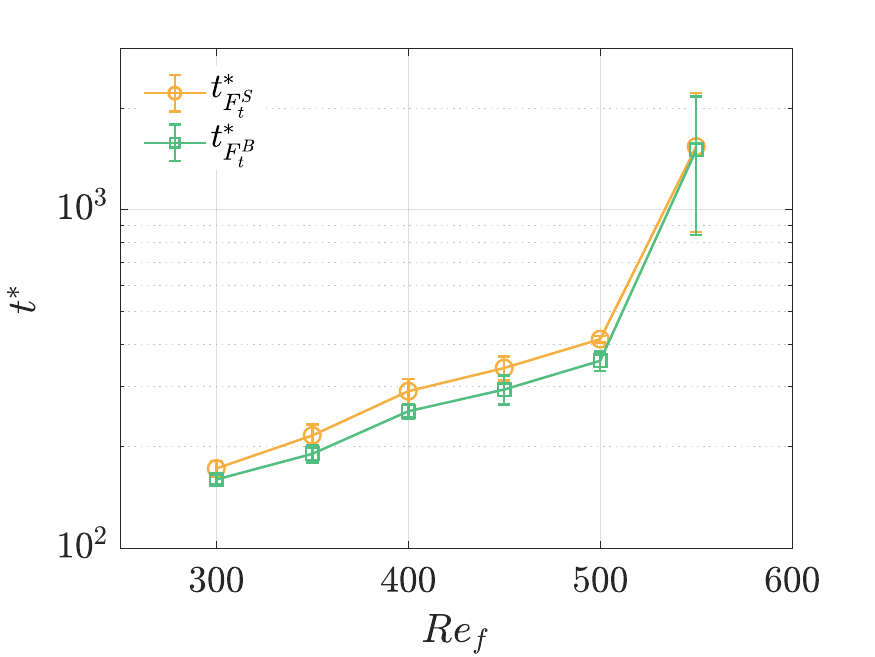}}
  \caption{Decay time $t^*$ as a function of the final Reynolds number $Re_f$, measured from the evolution of the turbulent fraction using two image processing methods: $t_{F_t^S}$ gold circles: the intensity-variation method \cite{Sano_NP_2016}, and $t_{F_t^B}$ green squares: the morphological method \cite{Bottin_thesis_1998}.}
\label{fig:decay_time_visulization}
\end{figure}
We compare the decay times $t^{*}$ obtained from the two
visualization-based turbulent-fraction measures, $F_t^{S}$ and $F_t^{B}$.
The corresponding nondimensionalized decay times, $t^{*}_{F_t^{S}}$ and
$t^{*}_{F_t^{B}}$ (scaled by $h/U_{\text{belt}}$), are plotted as functions of
$Re_f$ in Fig.~\ref{fig:decay_time_visulization}.  
Across the entire range $Re_f = 300$–$550$, the two estimates lie
virtually on top of each other, indicating that both image-processing
methods yield consistent decay times despite their differing algorithms
and sensitivities.  
This agreement confirms that the measurement of the streak-dominated
relaminarisation timescale is robust to the choice of visualization
processing method.

\begin{figure}[t!]
  \centerline{\includegraphics[width=10cm]{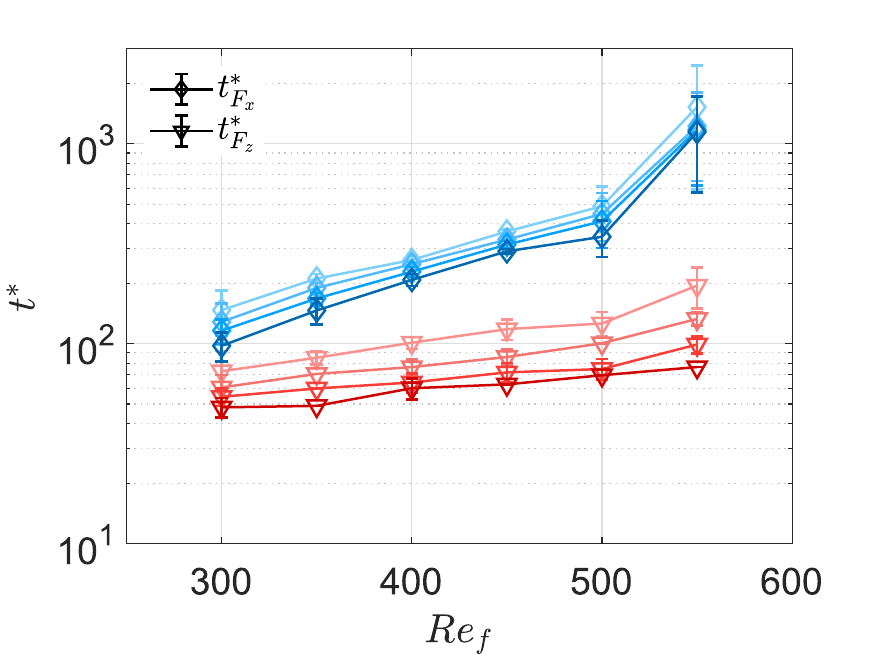}}
  \caption{Decay time $t^*$ as a function of the final Reynolds number $Re_f$, measured from the evolution of the turbulent fraction by thresholding the streamwise and spanwise velocities at different thresholds. Blue diamonds ($t_{F_x}$) correspond to thresholding the streamwise velocity with color variation from light to dark blue representing thresholds $|u_x| > 0.06$, 0.08, 0.10, and 0.12 $U_{\text{belt}}$. Red downward triangles ($t_{F_z}$) correspond to thresholding the spanwise velocity with color variation from light to dark red representing thresholds $|u_z| > 0.03$, 0.04, 0.05, and 0.06 $U_{\text{belt}}$.} 
\label{fig:decay_time_velocity_diff_thred}
\end{figure}

To compare the decay times obtained from image processing with those from
velocity–thresholding, we first examine the sensitivity of the latter to the
choice of threshold.  
Fig.~\ref{fig:decay_time_velocity_diff_thred} shows the nondimensional
decay times $t^{*}_{F_x}$ (blue diamonds) and $t^{*}_{F_z}$ (red downward
triangles) as functions of $Re_f$ for several streamwise and spanwise
velocity thresholds.  
For $F_x$, we vary the criterion $|u_x| > 0.06$, $0.08$, $0.10$, and
$0.12\,U_{\text{belt}}$; for $F_z$, we similarly vary $|u_z| > 0.03$,
$0.04$, $0.05$, and $0.06\,U_{\text{belt}}$, with marker color ranging
from dark to light as the threshold increases.

For both components, lowering the threshold systematically shifts the
estimated decay time upward, reflecting the longer persistence of weak
velocity fluctuations.  
However, the separation between $t^{*}_{F_x}$ and $t^{*}_{F_z}$ remains large
across all thresholds and all $Re_f$, confirming that the two components
retain distinct decay timescales.

For the subsequent comparisons, we adopt the thresholds
$|u_x| > 0.10\,U_{\text{belt}}$ and $|u_z| > 0.05\,U_{\text{belt}}$, which
provide a good compromise between capturing the morphology of the streaks and
suppressing spurious low-amplitude fluctuations, and which are consistent
with the criteria used in our previous study \cite{Liu_etal_2024}.

\begin{table}[!htbp]
  \centering
  \renewcommand{\arraystretch}{1.2}

    \scalebox{1}{
    \begin{tabular}{lcccccc}
      \hline
     Decay & \multicolumn{6}{c}{$Re_f$} \\
      \cline{2-7}
      time 
      & 300 & 350 & 400 & 450 & 500 & 550 \\
      \hline
      $t^{*}_{F_t^{S}}$ & $172.2\pm9.3$   & $215.6\pm16.6$ & $291.4\pm24.8$ & $341.2\pm27.9$   & $415.2\pm8.9$    & $1538.5\pm678.1$ \\
      $t^{*}_{F_t^{B}}$ & $159.8\pm5.4$   & $190.3\pm10.9$ & $254.2\pm12.4$ & $294.8\pm29.0$   & $358.5\pm23.6$   & $1504.5\pm661.9$ \\
      $t^{*}_{F_x}$     & $116.4\pm16.1$  & $168.5\pm10.9$ & $229.4\pm12.4$ & $313.4\pm13.9$   & $410.0\pm108.3$  & $1175.8\pm557.7$ \\
      $t^{*}_{F_z}$     & $54.3\pm5.4$    & $59.8\pm0$     & $64.1\pm7.2$   & $72.0\pm8.0$     & $74.8\pm8.9$     & $99.2\pm9.8$ \\
      $t^{*}_{E_x}$     & $97.7\pm16.1$   & $164.9\pm12.6$ & $233.5\pm14.3$ & $322.6\pm8.0$    & $435.8\pm103.0$  & $1232.5\pm579.2$ \\
      $t^{*}_{E_z}$     & $48.1\pm5.4$    & $56.2\pm6.3$   & $72.3\pm7.2$   & $85.9\pm8$       & $100.6\pm0$      & $144.5\pm17$ \\
      \hline
    \end{tabular}%
  }

  \caption{Decay times $t^{*}$ obtained from turbulent fraction and
kinetic energy evolutions for different final Reynolds numbers $Re_f$.  
These values correspond to the data shown in Fig.~\ref{fig:decay_time_visulization},~\ref{fig:decay_time_comparison_velocity_visu}.}
\label{tab:decay_time}
\end{table}

\begin{figure}[t!]
  \centerline{\includegraphics[width=10cm]{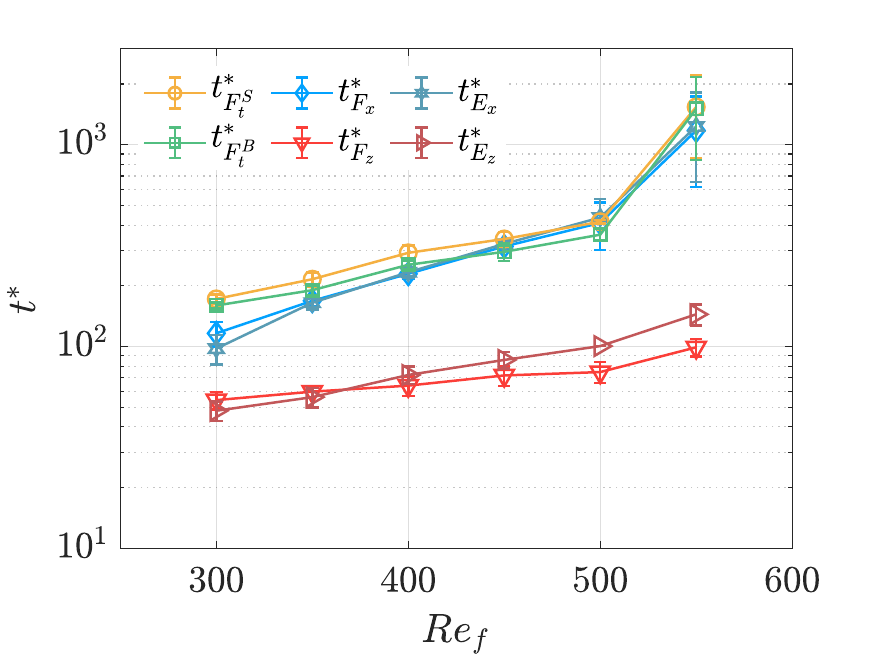}}
  \caption{Comparison of decay time $t^*$ obtained from flow visualization and PIV-based velocity thresholding as a function of the final Reynolds number $Re_f$.
Gold circles represent $t_{F_t^S}$ extracted using the intensity-variation method \cite{Sano_NP_2016}, and green squares show $t_{F_t^B}$ from the morphological method \cite{Bottin_thesis_1998}.
Blue diamonds correspond to threshold on the streamwise velocity $|u_x| > 0.1U_{\text{belt}}$, and red downward triangles to the spanwise velocity threshold $|u_z| > 0.05U_{\text{belt}}$.}
\label{fig:decay_time_comparison_velocity_visu}
\end{figure}

We now compare the decay times obtained from flow visualization, velocity thresholding, and the decay of the streamwise and spanwise kinetic energy, where the kinetic energy in each direction is defined as the mean square of the corresponding velocity component. Fig.~\ref{fig:decay_time_comparison_velocity_visu} summarizes the
nondimensional decay times extracted from the image–processing methods,
$t^{*}_{F_t^{S}}$ (yellow circles) and $t^{*}_{F_t^{B}}$ (green squares); the
velocity–threshold measures, $t^{*}_{F_x}$ (blue diamonds) and
$t^{*}_{F_z}$ (red downward triangles); and the kinetic energy-based quantities,
$t^{*}_{E_x}$ (dark–blue hexagrams) and $t^{*}_{E_z}$ (dark–red
right-pointing triangles), over the range of final Reynolds numbers
$Re_f$. The detail of the decay time is listed in Table~\ref{tab:decay_time}.

Across all $Re_f$, the decay times obtained from visualization agree closely
with the streamwise measures $t^{*}_{F_x}$ and $t^{*}_{E_x}$, and remain
substantially larger than the spanwise quantities $t^{*}_{F_z}$ and
$ t^{*}_{E_z}$.  
This consistent ordering reflects the separation of streak and roll
timescales: visualization predominantly tracks the slow decay of the
streamwise streaks, whereas the spanwise-roll dynamics decay much more
rapidly.  
Despite their differing definitions, all three approaches visualization, velocity thresholding, and energy decay 
yield compatible streamwise decay times, indicating that each method 
reliably captures the evolution of the streak-dominated 
relaminarisation process.  
The slightly longer decay times obtained from intensity-based 
visualization arise because this method retains small, isolated 
turbulent regions that remain visible even when their velocity 
fluctuations are very weak.  
In contrast, the morphological filtering used in the visualization thresholding approach removes structures below prescribed width and length scales through image opening operations, prioritizing the observation of straight streaks.

These results reinforce that rheoscopic flow visualization remains a powerful
diagnostic for studying transitional wall-bounded flows, especially when
optical constraints make PIV difficult to implement.  
Recent work in pipe flow~\cite{Raj_2025} has shown that flow
regimes can be identified from the evolution of streak orientations,
highlighting the broader potential of visualization-based diagnostics for
tracking the spatiotemporal organization of turbulence pipe flows.

\section{Conclusions}
\label{sec:Conclusion}

We have experimentally investigated what rheoscopic flow visualization measures in plane Couette–Poiseuille flow by performing quench experiments and comparing the results directly with PIV velocity fields.
Decay times were extracted in three ways: from visualization-based turbulent fractions using two image-processing methods inspired by \cite{Bottin_thesis_1998,Sano_NP_2016}, and from PIV using both velocity thresholding and the decay of the streamwise and spanwise kinetic energy.
Across the range of final Reynolds numbers explored, the decay times obtained from visualization are consistent with those associated with the streamwise velocity component and are significantly larger than those associated with the spanwise component.
This demonstrates that rheoscopic visualization primarily tracks the decay of streamwise streaks rather than the faster decay of the rolls.

Two visualization-processing methods were examined.
The intensity-variation method of \cite{Sano_NP_2016} and the morphological method of \cite{Bottin_thesis_1998} both produce robust turbulent-fraction evolutions and yield decay times in good agreement with the PIV-based streamwise measures.
A transient increase in the visualization-based turbulent fraction is consistently observed immediately after the quench. Streak broadening occurs during the early stages of relaminarization: residual streaks weaken in amplitude while expanding laterally, leading the image-processing algorithms to classify a larger region as turbulent before the remaining shear eventually falls below the detection threshold. In contrast, the fully turbulent state exhibits finer, more fractured streaks whose morphology is dominated by undulations linked to normal vorticity. 

Among the two visualization methods, the intensity-variation approach preserves streak morphology most faithfully and resolves features such as waviness and breakup that can be difficult to capture with PIV at comparable spatial resolution.
The morphological method produces smoother contours but still captures the large-scale organization and yields similar decay times.
Consistent with earlier findings \cite{Savaş_1985,Gauthier_1998,Weidman_1976,Carlson_1982}, our results confirm that the reflected intensity in rheoscopic images arises from the orientation of thin flakes in response to local shear and velocity-gradient structures.
This sensitivity allows visualization to detect weak shear regions that remain below the velocity thresholds used in PIV-based diagnostics.

Recent work has shown that long streamwise straight streaks are not required for the self-sustenance of the bursting process in wall-bounded turbulence and may instead arise as by-products of shorter scale dynamics, with the self regeneration mechanism persisting in their absence \citep{Liu_etal_2024,Etchevest2025,Jimenez_2022}. In this context, particle image velocimetry directly measures the spanwise vorticity of the roll component in the self-sustaining cycle. Rheoscopic visualization, however, predominantly reveals streak modulations—including both the intense streaks involved in regeneration and weaker, elongated streaks whose immediate dynamical contribution is less direct. A perspective would be to extract the streak waviness, which is another essential component of the self-sustaining process \citep{Liu_etal_2024,Etchevest2025}.

Overall, the comparison between visualization-based turbulent fractions, PIV velocity thresholding, and kinetic-energy decay clarifies the physical content of rheoscopic measurements in wall-bounded shear flows.
Flow visualization, when combined with appropriate image-processing techniques, provides a reliable measure of streak persistence during relaminarization and serves as a valuable complementary diagnostic.

\section{Acknowledgment}

We acknowledge Amaury Fourgeaud, Olivier Brouard, Xavier Benoit-Gonin, and Thierry Darnige from the PMMH for their patient and invaluable help with the experimental setup, and Manuel Etchevest and Matías Ramdan Ferressini for fruitful discussions.

\bibliography{Reference}

\end{document}